\documentclass[12pt]{JHEP3}
\usepackage{amsmath,amssymb,epsfig}

\font\cmss=cmss12

\newcommand{\bc}{\begin{center}}
\newcommand{\ec}{\end{center}}
\newcommand{\ba}{\begin{array}}
\newcommand{\ea}{\end{array}}
\newcommand{\beq}{\begin{equation}}
\newcommand{\eeq}{\end{equation}}
\newcommand{\bea}{\begin{eqnarray}}
\newcommand{\eea}{\end{eqnarray}}
\newcommand{\bmx}{\begin{pmatrix}}
\newcommand{\emx}{\end{pmatrix}}
\newcommand{\nn}{\nonumber}

\newcommand{\half}{\frac{1}{2}}

\newcommand{\taubar}{{\overline \tau}}

\newcommand{\tp}{{\tilde p}}

\newcommand{\eref}[1]{Eq.~(\ref{#1})}

\newcommand{\cE}{{\cal E}}

\newcommand{\vQ}{{\vec{Q}}}
\newcommand{\vP}{{\vec{P}}}

\newcommand{\vN}{{\vec{N}}}
\newcommand{\vp}{{\vec{p}\,}}
\newcommand{\vk}{{\vec{k}\,}}

\newcommand{\bps}{{\rm BPS}}

\def\IB{\relax{\rm I\kern-.18em B}}
\def\IC{{\relax\hbox{\kern.3em{\cmss I}$\kern-.4em{\rm C}$}}}
\def\ID{\relax{\rm I\kern-.18em D}}
\def\IE{\relax{\rm I\kern-.18em E}}
\def\IF{\relax{\rm I\kern-.18em F}}
\def\II{\relax{\rm I\kern-.18em I}}
\def\IZ{\relax{\sf Z\kern-.35em Z}}
\def\Id{\relax{1\kern-.32em 1}}
\def\IG{\relax\hbox{$\inbar\kern-.3em{\rm G}$}}
\def\IR{\relax{\rm I\kern-.18em R}}

\normalsize

\renewcommand{\Re}{{\rm Re\,}}

\title{Kinematical Analogy for Marginal Dyon Decay} 
\author{Anindya Mukherjee\,\footnote{Email:
    anindya\_m@theory.tifr.res.in}\,, Sunil Mukhi\,\footnote{Email:
    mukhi@tifr.res.in}\,, and Rahul Nigam\,\footnote{Email:
    rahulnig@theory.tifr.res.in}\\ \it Tata Institute of Fundamental
    Research,\\ \it Homi Bhabha Rd, Mumbai 400 005, India}

\preprint{TIFR/TH/07-31}


\abstract{We describe a kinematical analogy for the
marginal decay of $\frac14$-BPS dyons in 4-dimensional ${\cal N}=4$
string compactifications. In this analogy, the electric and magnetic
charges play the role of spatial momenta, the BPS mass plays the role
of energy, and $\half$-BPS dyons correspond to massless
particles. Using $SO(12,1)$ ``Lorentz'' invariance and standard
kinematical formulae in particle physics, we provide simple
derivations of the curves of marginal stability. We also show how
these curves map into the momentum ellipsoid, and propose some
applications of this analogy.}

\begin{document}

\subsection*{Introduction and review}

Dyonic black holes in type IIB superstring compactifications to four
dimensions on $K3\times T^2$ have been the subject of much interest in
recent times. The counting formula of Ref.\cite{Dijkgraaf:1996it} for
the microstates of these black holes has been re-derived in different
ways\cite{Shih:2005uc,Gaiotto:2005hc,David:2006yn} and subtle jumps in
the spectrum have been accounted for by invoking dyon decay across
curves of marginal stability\cite{Dabholkar:2006bj,%
Sen:2007vb,Dabholkar:2007vk,Sen:2007pg,%
Cheng:2007ch,Sen:2007nz,Mukherjee:2007nc}.

Marginal stability is understood in terms of a formula that expresses
the BPS mass of a dyonic state in terms of electric and magnetic
charges as well as the moduli of the compactification. The BPS mass
formula in turn descends from the supersymmetry algebra in 10
dimensions. For BPS states, only the right-moving momenta (the 6 part
of $(6,22)$) are relevant, so one expects to see an $SO(6)$ symmetry.
However, the phenomenon of interest to us, namely marginal decay of
dyons, involves a larger symmetry. In addition to the 6 relevant
electric charges there are now 6 magnetic charges. As we will see, the
formulae relevant to dyon decay turn out to have an $SO(12,1)$
invariant structure that does not inherit from a symmetry of the
underlying string theory. We will exploit this symmetry to recast the
kinematics of dyon decay in terms of particles propagating in $12+1$
dimensions. This allows us to use standard formulae from particle
kinematics to study marginal dyon decay.

The system we work with is the ${\cal N}=4$ supersymmetric background
in 4 spacetime dimensions corresponding to type IIA/IIB string theory
compactified on $K3\times T^2$ or the heterotic string on $T^6$, the
two being related by duality. In what follows we will use the duality
frame appropriate for type IIB. The resulting four-dimensional system
has 28 $U(1)$ gauge fields along with their electric-magnetic duals,
therefore we can have dyons of electric and magnetic charge $\vQ$ and
$\vP$, each one being a 28-vector. These dyons will be $\half$-BPS if
$\vQ,\vP$ are parallel, and $\frac14$-BPS otherwise.

The BPS mass formula depends on the charges carried by the dyons as
well as the values of the compactification moduli.  These are encoded
as follows. Define the matrix:
\beq
L\equiv {\rm diag}(1^6;(-1)^{22})
\eeq
There are 132 moduli that can be assembled into a symmetric matrix $M$
which is orthogonal with respect to the $L$ metric:
\beq
M^T=M,\quad M^T LM = L
\eeq
For purposes of the BPS mass formula, the relevant inner products for
the charge vectors are:
\beq
\label{moddepprod}
Q^2 \equiv \vQ^T(M+L) \, \vQ,\qquad
P^2 \equiv \vP^T(M+L)\,\vP,\qquad 
P\cdot Q \equiv \vP^T(M+L)\,\vQ
\eeq
Let us also define the right-moving projections of the charge vectors,
$\vQ_R,\vP_R$, by:
\beq 
\label{projected}
Q_R^2\equiv\vQ_R^T\vQ_R=\vQ^T(M+L) \, \vQ =Q^2
\eeq 
and similarly for the other inner products. In this way the
moduli-dependent inner product \eref{moddepprod} for the charge
vectors is replaced by a standard moduli-independent product while the
moduli-dependence is moved to the vectors themselves\footnote{This
refers to all moduli other than those of the 2-torus, which are always
displayed explicitly and do not appear in the inner product.}. In what
follows, we will not always write the suffix $R$, since the inner
products are by definition the same whether we are dealing with the
projected or unprojected vector.

The BPS mass formula for general $\frac14$-BPS dyons is
\cite{Cvetic:1995uj,Duff:1995sm,Sen:2007vb}:
\beq
\label{genbpsmass}
M^\bps(\vQ,\vP)^2 =
\frac{1}{\tau_2}(\vQ-{\bar\tau}\vP)\cdot(\vQ-{\tau}\vP) +
2\sqrt{\Delta(\vQ,\vP)} \eeq
where
\beq
\label{Deltadef}
\Delta(\vQ,\vP) \equiv Q^2 P^2 - (P\cdot Q)^2
\eeq
and $\tau$ is the modular parameter of the torus\footnote{Our
conventions are those in Ref.\cite{Sen:2007vb} and differ by a factor
of $\frac{1}{\sqrt{\tau_2}}$ from those in Ref\cite{Mukherjee:2007nc}.}.

\subsection*{Kinematic Analogy}

The formula \eref{genbpsmass} has a striking analogy to the
energy-momentum dispersion relation for a relativistic point
particle, $E^2 = \vp^{2} + m^2$. In this analogy, the BPS mass plays the
role of the energy of the particle, while the dyonic charges play the
role of momenta:
\beq
\label{momchargeinit}
\vp = \frac{1}{\sqrt{\tau_2}}(\vQ - \tau\vP)
\eeq
Conservation of dyonic charge corresponds to momentum conservation in
the kinematical system. For this, it is crucial that $\tau$ be complex
(in fact it has a strictly positive imaginary part). Then conservation
of the imaginary part of the momentum is equivalent to conservation of
the dyon's magnetic charge, while the real part ensures conservation
of electric charge.

To complete the analogy, note that the condition of marginal stability
in a decay: $M^\bps = M_1^\bps + M_2^\bps+\cdots$ corresponds to
energy conservation. Finally, the mass of the analogue particle is
$\sqrt2\left(\Delta(\vQ,\vP)\right)^{\frac14}$. A $\half$-BPS dyon has
its electric and magnetic charge vectors proportional, so
$\Delta(\vQ,\vP)=0$. It therefore corresponds to a massless particle.

Because the momentum vector defined in \eref{momchargeinit} is
complex, some care must be taken in defining the inner product. As
each of $\vQ,\vP$ is effectively 6-dimensional (recall that they are
all projected using a moduli-dependent matrix, as in
\eref{projected}), one can convert the momenta to
real vectors in a 12-dimensional space with Lorentz group
$SO(12,1)$. As we have noted, these Lorentz symmetries are symmetries
of the dispersion relation and the kinematics of the decay process,
but not of the underlying string theory.

This point can be clarified by considering a general dyonic particle
having a momentum $\vk$ in the noncompact three-dimensional space. For
such a particle the full dispersion relation is:
\bea
E^2 &=& \vk^2 + (M^{\bps})^2\nn\\
&=& \vk^2 + \frac{1}{\tau_2}(\vQ-{\bar\tau}\vP)\cdot(\vQ-{\tau}\vP) +
2\sqrt{\Delta(\vQ,\vP)}
\eea
If we restrict our attention to electrically charged particles,
$\vP=0$, the above equation simplifies to:
\beq
E^2 = \vk^2 + \frac{1}{\tau_2}\vQ^2
\eeq
which can be thought of as the dispersion relation for either a
massive particle in 4d or a massless particle in 10d. This is the
expected situation for a Kaluza-Klein particle whose electric charges
are momenta along the toroidal directions. The dependence on the
modulus $\tau$, though present, is trivial in this case and can be
removed by rescaling the metric appropriately. However, once we
introduce magnetic charges as well then the dispersion relation is
nontrivially $\tau$-dependent and also contains the $\Delta$
factor. Now if we put $\vk=0$ then we have the kinematic analogy of
interest in the present work.

To recover the curves of marginal stability, consider a general
two-body decay of a particle with energy-momentum $(E,\vp)$ into a
pair of particles of energy-momentum $(E_1,\vp_1)$ and
$(E_2,\vp_2)$. There are two Lorentz frames that are useful: the rest
frame of the initial particle and the lab frame where the initial
particle has a given spatial momentum. By working in the rest frame,
the Lorentz invariant $p\cdot p_1$ is easily shown to be:
\beq
p\cdot p_1 = \half(m^2+m_1^2-m_2^2)
\eeq
The same Lorentz invariant in the lab frame can be written:
\beq
p\cdot p_1 = m_1^2 + \sqrt{\vp_1^{2} + m_1^2} \sqrt{\vp_2^{2} + m_2^2} - 
\vp_1\cdot \vp_2
\eeq
Equating the two, we have:
\beq
\label{momeqnone}
\sqrt{\vp_1^{2}+m_1^2}\sqrt{\vp_2^{2} + m_2^2} - \vp_1\cdot \vp_2
= \frac{m^2-m_1^2-m_2^2}{2}
\eeq

To apply this to the decay of a $\frac14$-BPS dyon, we make the
substitution in \eref{momchargeinit} as well as:
\beq
\label{momchargefin} 
\vp_1 = \frac{1}{\sqrt{\tau_2}}(\vQ_1 - \tau\vP_1),
\quad
\vp_2 = \frac{1}{\sqrt{\tau_2}}(\vQ_2 - \tau\vP_2)
\eeq
to find:
\bea
\label{gencurve}
\sqrt{|\vQ_1-\tau\vP_1|^2+ 2\tau_2\sqrt\Delta_1}
\sqrt{|\vQ_2-\tau\vP_2|^2+ 2\tau_2\sqrt\Delta_2}
&-& {\rm Re} (\vQ_1-\tau\vP_1)\cdot(\vQ_2-\taubar\vP_2)\nn\\
&=& \tau_2(\sqrt\Delta - \sqrt\Delta_1 - \sqrt\Delta_2)
\eea
where $\Delta_i = \Delta(\vQ_i,\vP_i)$. For fixed charge vectors
$\vQ_i,\vP_i$ and moduli matrix $M$ this is an equation for the torus
modular parameter $\tau$. In other words, this is the general curve
of marginal stability!

We can relate this to previously derived forms of the curve, as in
\cite{Sen:2007vb,Sen:2007nz,Mukherjee:2007nc}. First let us
consider the case of decay of a primitive ($gcd(\vQ\wedge
\vP)=1$) $\frac14$-BPS dyon into two $\half$-BPS dyons. In this case the
product particles are massless. Moreover, the charge vectors of the
decay products are given by\cite{Sen:2007vb}:
\bea
\vQ_1 &=& a\vN_1,\quad \vP_1 = c\vN_1 \nn\\
\vQ_2 &=& b\vN_2,\quad \vP_2 = d\vN_2
\eea
where
\beq
\vN_1= d\vQ-b\vP,\quad \vN_2 = -c\vQ + a\vP
\eeq
nd $ad-bc=1$. Thus the equation reduces to:
\beq
|a-c\tau||b-d\taubar|\,\frac{|\vN_1| |\vN_2|}{\sqrt\Delta}
- {\rm Re}\left\{(a-c\tau)(b-d\taubar)\right\}
\frac{\vN_1\cdot\vN_2}{\sqrt\Delta} =
\tau_2
\eeq
Let
\beq
\label{calEdef}
\cE = -\frac{\vN_1\cdot \vN_2}{\sqrt{\Delta(\vN_1,\vN_2)}}
\eeq
It is easily checked that
\beq
\sqrt{1+\cE^2} = \frac{|\vN_1||\vN_2|}{\sqrt{\Delta(\vN_1,\vN_2)}}
\eeq
Using the above equations and the fact that
$\Delta(\vN_1,\vN_2)=\Delta(\vQ,\vP)$, the curve becomes:
\beq
|a-c\tau||b-d\taubar|\,\sqrt{1+\cE^2} + 
{\rm Re}\left\{(a-c\tau)(b-d\taubar)\right\}\cE =\tau_2
\eeq
Now one can take the second term to the right side and square both
sides, whereupon the resulting quartic (in $\tau$) re-factorises into
a perfect square of the familiar Sen circle equation. However there is
a neater way to proceed. The above equation is the same as:
\beq
|(a-c\tau)(b-d\taubar)(\cE+i)| + {\rm Re}(a-c\tau)(b-d\taubar)(\cE+i)=0
\eeq
which is equivalent to the two conditions:
\bea
{\rm Im}(a-c\tau)(b-d\taubar)(\cE+i) &=& 0\nn\\
{\rm Re}(a-c\tau)(b-d\taubar)(\cE+i) &<& 0
\eea
The first of these is directly the equation of the Sen circle:
\beq
\left(\tau_1-\frac{ad+bc}{2cd}\right)^2 + 
\left(\tau_2 + \frac{\cE}{2cd}\right)^2 =
\frac{1}{4c^2d^2}(1+\cE^2)
\eeq
while the second one restricts us to the $\tau_2>0$ region of that
circle.

Next consider decays into two $\frac14$-BPS dyons, or one
$\frac14$-BPS and one $\half$-BPS dyon. In this case the charges of
the final states are parametrised as:
\bea
\vQ_1 = m_1\vQ+r_1\vP,\quad &&\vP_1 = s_1\vQ+n_1\vP \nn\\
\vQ_2 = (m-m_1)\vQ - r_1\vP,\quad &&\vP_2 = -s_1\vQ + (n-n_1)\vP
\eea
where the charges of the initial dyon are now $(m\vQ,n\vP)$ with
$gcd(\vQ\wedge\vP) =1$.

Substituting the above into \eref{gencurve}, transposing the second
term to the right hand side and squaring, one finds:
\beq
\label{cms}
\left(\tau_1-\frac{m\wedge n}{2ns_1}\right)^2
+ \left(\tau_2 + \frac{\cE}{2ns_1}\right)^2
= \frac{1}{4n^2s_1^2}\Big((m\wedge n)^2 + 4mnr_1s_1 + \cE^2\Big)
\eeq
where
\beq
\cE\equiv -\frac{\vQ_1\cdot \vP_2 - \vQ_2\cdot \vP_1}{\sqrt{\Delta}}=
\frac{1}{\sqrt\Delta}\left(ms_1\, Q^2 -nr_1\, 
P^2 - (m\wedge n)Q\cdot P\right)
\eeq
and $m\wedge n = m_1 n_2 - m_2 n_1$. This is the general curve of
marginal stability found in \cite{Mukherjee:2007nc}.

\subsection*{Momentum ellipsoid}

For two-body decay of an unstable particle in the lab frame, the
final-state particle momenta are constrained to lie on an ellipsoid of
revolution, obtained by rotating an ellipse in the forward and
transverse momenta $p_{1,\parallel}$ and $p_{1,\perp}$ along the
azimuthal angles around the beam axis. It is tempting to guess that
the curve of marginal stability coincides with this ellipse, We will
see below that this is roughly true but the relationship is more
complicated than one might have expected.

Indeed this approach, in which $p$ and $p_1$ are treated as the
independent variables, is not the easiest way to {\it derive} the
curves of marginal stability, which in fact we have already done in
the previous section by treating the final-state momenta $p_1,p_2$ as
the independent variables.  Nevertheless it is of some conceptual
interest to understand how the curves of marginal stability are
related to the momentum ellipsoid. One reason why the embedding we
will obtain is rather complicated is that the momentum of the
initial particle (which determines one axis of the ellipse) itself
depends on the modular parameter $\tau$.

To find the momentum ellipsoid, we first eliminate $\vp_2$ in
\eref{momeqnone} in favour of $\vp,\vp_1$. Then, taking the second
term to the right hand side and squaring, we end up with the equation:
\beq
\label{momellipse}
m^2\left(p_{1,\parallel} - \frac{|\vp|(m^2-m_1^2-m_2^2)}{2m^2}\right)^2
+ (\vp^2+m^2)\,p_{1,\perp}^2 = \frac14
\lambda(m^2,m_1^2,m_2^2)\left(1+\frac{\vp^2}{m^2}\right) 
\eeq
where
\beq
\lambda(a,b,c)\equiv a^2 + b^2 + c^2 - 2ab -2bc -2ca
\eeq
Here $p_{1,\parallel}$ and $p_{1,\perp}$ are the components of
$\vp_{\!1}$ along and transverse to the beam.  The momentum ellipsoid
is the ellipse in \eref{momellipse} rotated about the beam axis.

For decays into two $\half$-BPS states, we put $m_1=m_2=0$ and the
momentum ellipsoid simplifies considerably into:
\beq
\frac{\Big(p_{1,\parallel}-\frac{|\vp|}{2}\Big)^2}{\frac14(\vp^2+m^2)} +
\frac{p_{1,\perp}^2}{\frac14 m^2}=1
\eeq
Evidently the major axis of the ellipse is proportional to
$\sqrt{\vp^2+m^2}$, which from \eref{momchargeinit} is
$\tau$-dependent. Therefore the momentum ellipsoid itself varies with
$\tau$, and the curve of marginal stability is technically not a
subspace of a particular ellipsoid. This can be remedied by defining a
new variable $\tp_1$ via:
\beq
\tp_1 \equiv
\frac{p_{1,\parallel}-\frac{|\vp|}{2}}{\sqrt{1+\frac{\vp^2}{m^2}}}
\eeq
in terms of which the ellipse becomes a circle with a
$\tau$-independent radius:
\beq
\tp_1^2
+p_{1,\perp}^2
= {\frac14 m^2}
\eeq

From Eqs.(\ref{momchargeinit}),(\ref{momchargefin}) we find:
\bea
p_{1,\parallel} &=& \frac{1}{\sqrt{\tau_2}}
\frac{\Re\Big( (\vQ-\tau\vP)\cdot (\vQ_1-\taubar\vP_1)
\Big)}{|\vQ-\tau\vP|}\\
p_{1,\perp} &=& \frac{1}{\sqrt{\tau_2}}
\frac{1}{|\vQ-\tau\vP|}
\sqrt{|\vQ-\tau\vP|^2 |\vQ_1-\tau\vP_1|^2 - 
\Big(\Re(\vQ-\tau\vP)\cdot(\vQ_1-\taubar\vP_1)\Big)^2}\nn
\eea
from which one obtains:
\bea
\tp_1 &=& \frac{1}{\sqrt{\tau_2}}\frac{1}{|\vQ-\tau\vP|}
\frac{(ad+bc-2cd\tau_1)Q^2 + (2ab\tau_1 - (ad+bc)|\tau|^2)P^2
+ (-2ab+2cd|\tau|^2) Q\cdot P}{\sqrt{\frac{1}{\tau_2}|\vQ-\tau\vP|^2 +
2 \sqrt{\Delta(\vQ,\vP)}}}\nn\\
p_{1,\perp} &=&
\frac{1}{\sqrt{\tau_2}}\frac{1}{|\vQ-\tau\vP|}\sqrt{\Delta(\vQ,\vP)} 
\sqrt{\Big(\Re(a-c\tau)(b-d\taubar)\Big)^2 + (\cE^2+1)\tau_2^2}
\eea
where $\cE$ has been defined in \eref{calEdef}. As promised, this is
rather complicated. One might hope to find a better parametrisation
that leads to a simpler embedding.

\subsection*{Multiparticle decays and the issue of codimension}

In order for the dyons in the final state to be BPS relative to each
other, $\vQ_{i,R},\vP_{i,R}$ must lie in the same plane as
$\vQ_R,\vP_R$\footnote{This condition also follows from the
kinematics\cite{Sen:2007nz}, we thank the referee for emphasizing this
fact to us.}. This condition implies that all decays other than two-body
decays into $\half$-BPS particles occur on codimension $\ge 2$
subspaces of moduli space, as noted in
Refs.\cite{Sen:2007nz,Mukherjee:2007nc}.

We can incorporate this property into the kinematical analogy. Let us
impose on our 12-dimensional space the structure of a linear
symplectic manifold. Thus there is a polarisation - a closed,
non-degenerate 2-form that divides the space into 6 electric and 6
magnetic directions, each one a Lagrangian subspace analogous to
coordinates and momenta in a phase space.

Consider a particle of arbitrary 12-momentum. We use the polarisation
to divide this momentum into its electric and magnetic parts, each
being a 6-vector. Now consider the ``diagonal'' 6-manifold obtained by
identifying the electric and magnetic subspaces. In this 6-manifold,
the linear span of the electric and magnetic parts of the 12-momentum
defines a plane (if these 6-vectors are non-parallel) or a line (if
they are parallel). The former case is $\frac14$-BPS while the latter
is $\frac12$-BPS.

In a two-body decay, if the original particle defines a plane in the
``diagonal'' space while the final particles define lines, then by
momentum conservation the BPS condition is automatically satisfied. If
the final state particles are $\half$-BPS but three or more in number,
or if at least one of them is $\frac14$-BPS, then the dimensionality
of the subspace of the diagonal space spanned is at least three. In
this case, additional conditions on the moduli besides the marginal
stability condition are required to make the decay
possible\cite{Sen:2007nz,Mukherjee:2007nc}. The embedding of
codimension $\ge 2$ curves in the full moduli space has, however, not
yet been given a precise description.

\subsection*{Decay widths on marginal stability curves}

The kinematical analogy suggests that one consider decay and
scattering processes involving dyons on curves of marginal
stability. As a simple example, consider an ensemble of $\frac14$-BPS
dyons of a given charge, at a curve of marginal stability for decay
into two $\half$-BPS dyons. The ensemble will decay with a width given
by a formula analogous to the classic formula for decay of a particle
of mass $m$ into two identical massless particles, which in three
noncompact space dimensions and in the rest frame of the decaying
particle is (see for example Ref.\cite{Griffiths:1987tj}):
\beq
\Gamma = \frac{1}{32\pi\hbar m}|{\cal M}|^2
\eeq
where ${\cal M}$ is the matrix element for the process.

To apply this formula to the present case, the kinematics needs to be
re-done in the lab frame of the decaying particle, and in 12
dimensions. Also we need to incorporate the true spacetime momenta of
the initial and final state particles - the above expression should be
thought of only as the factor in the decay width coming from the
``analogue'' momenta. We also need to take into account the
quantisation rule for the ``analogue'' 12d momenta, inherited from the
quantisation of the original electric and magnetic charges. Finally,
the matrix element ${\cal M}$ needs to be computed. It seems quite
plausible that everything here is computable in string theory. One can
similarly consider scattering cross-sections for dyons.

\subsection*{Conclusions}

We have presented an analogy that maps the marginal stability
conditions for $\frac14$-BPS dyons into energy-momentum conservation
in an analogue particle problem in 12+1 dimensions. $\frac14$-BPS
states behave like massive particles and $\frac12$-BPS states like
massless particles. The analogy provides a simple way to understand
curves of marginal stability and may be useful both in deriving these
curves in other situations and in suggesting ways to think about
physical processes involving $\frac14$-BPS dyons.

\subsection*{Acknowledgements}

We would like to thank Rajesh Gopakumar and Sandip Trivedi, as well as
all the participants of the Indian Strings Meeting, HRI Allahabad,
October 2007, for useful discussions. The work of AM was supported in
part by CSIR Award No. 9/9/256(SPM-5)/2K2/EMR-I. As always, the
generous support of the people of India is gratefully acknowledged.

\bibliographystyle{JHEP}

\bibliography{kin-analogy}

\end{document}